\documentstyle[12pt,twoside]{article}
\begin{document}
\title{\begin{flushright}
UMD-PP-94-141\\
May 1994 \ \ \\
\end{flushright}
CP-like Symmetry, Family Replication, \\
Charge Quantization}
\author{Ravi Kuchimanchi  \\
\vspace{.1in} \\
Department of Physics, University of Maryland\\
College Park, Maryland 20742\\
E-mail:  ravi@umdhep.umd.edu\\
Bulletin Board: hep-ph/9405276}
\date{}
\maketitle

\begin{abstract}
We propose that the physics beyond the standard Weinberg-Salam
model is such that
matter and the CP conjugate anti-matter fields have the same set
of charges with respect to the various force groups (upto ordering).
We show that this CP-like symmetry leads to some understanding of family
replication.  We use anomaly constraints to infer the fermion spectra
that lead to automatic electric charge quantization.  We also find that
the unification of coupling constants is possible at $M_{U} = 2 \times
10^{15} GeV$.
\end{abstract}
\newpage
It is paradoxical that we have many good reasons to believe that
many particles not yet discovered actually exist, but hardly any
reasons to justify  the existence of some of the particles
already discovered!
Despite lot of effort and progress in particle physics, there
have been very few suggestions
and ideas as to why nature has picked three families of quarks and
leptons, instead of just one family.
Neither grandunified theories nor left-right symmetries provide an
answer to this question. No definite answer or prediction has come out
of supersymmtery or superstring theories.
Horizontal symmetries \cite{hor,yan,kan,zou} have been proposed between
the three families, but the choice of the horizontal group
has been largely arbitrary.
They provide no explanation to the family
triplication puzzle, since the horizontal group is
guessed by looking at the families. So far there has been  no
independent way of guessing this group.
There are very few ideas like
the supersymmetric preon model \cite{pat}
and the $SU(3)_{L}\times SU(3)_{c} \times
U(1)_{X}$ model \cite{fra}
which provide possible solutions.
Since there is such a dearth of solutions, to what is one of the most
intriguing puzzles of the standard model, it is useful to investigate
ideas that might throw some light on this problem.  In this letter
we find that a CP-like discrete symmetry is useful to understand this
problem.

In the standard Weinberg-Salam model \cite{wei},
the $SU(2)_{w} \times SU(3)_{c} \times U(1)_{Y}$ assignments of the quarks
and leptons and their CP-conjugate states are as follows:
\begin{equation}
Q_{L} (2,3,1/3) \stackrel{CP}{\leftrightarrow}
Q^{c}_{R} (2,\bar{3},-1/3),
\end{equation}
$$\ u_{R} (1,3,4/3) \leftrightarrow u^{c}_{L} (1,\bar{3},-4/3), \ d_{R}
(1,3,- 2/3) \leftrightarrow d^{c}_{L} (1,\bar{3},2/3),$$
$$L_{L} (2,1,-1) \leftrightarrow L^{c}_{R} (2,1,1), \ e_{R} (1,1,-2)
\leftrightarrow e^{c}_{L} (1,1,2)$$
The interesting thing is that the $SU(2)$ quantum numbers
for the  left(right)-handed states  and their corresponding
right(left)-handed
anti-states is exactly the same for all the above fermions.  This
is because $\bar{2}$ and $2$ are equivalent
representations \cite{geo} of $SU(2)$.  However the $U(1)_Y$ and the
$SU(3)_c$ quantum numbers \cite{ft1}
are not the same for the two CP conjugate
states of the fermions.  We propose to "symmetrize the charges" of the
left-handed particle and right-handed anti-particle states,
by suggesting that there are two new forces, $U(1)_{new Y}$ and
$SU(3)_{new C}$.  If any left-handed fermion is assigned to the
complex representation
$(R,y)$ of the usual $SU(3)_{c} \times U(1)_{Y}$ then
its CP conjugate state, namely, the right-handed anti-fermion will
be assigned to the {\it same} representation $(R,y)$ of $SU(3)_{new c}
\times U(1)_{new Y}$.  Further, the gauge coupling constants
$g_{y} = g_{new y}$ and $g_{c} = g_{new c}$ so that
the action is invariant under the following operation
(which we will call Strong Parity or S Parity):
$$ S \ Parity:  \psi_{L}({\bf x},t)  \rightarrow
CP(\psi_{L}) \equiv \psi^{c}_{R}({\bf -x},t), \ \ \psi_{R}({\bf x},t)
\rightarrow CP(\psi_{R}) \equiv \psi^{c}_{L}({\bf -x},t) ,$$
\begin{equation}
SU(3)_{c} \leftrightarrow
SU(3)_{new c}, \ \ U(1)_{Y} \leftrightarrow U(1)_{new Y}, \ \  SU(2)_{w}
\leftrightarrow SU(2)_{w}
\end{equation}
where by $SU(3)_{c} \leftrightarrow SU(3)_{new c}$ we mean that
$G_{\mu}^{a}({\bf x},t)  \leftrightarrow H^{\mu a}({\bf -x},t) $, by
$U(1)_{Y} \leftrightarrow U(1)_{new Y}$ we  mean $A_{\mu}({\bf x},t)
\leftrightarrow
B^{\mu}({\bf -x},t) $, and by $SU(2)_{w} \leftrightarrow SU(2)_{w}$ we
mean $W_{\mu}^{a}({\bf x},t)  \leftrightarrow W^{\mu a}({\bf -x},t) $.
$G_{\mu}^{a}, H_{\mu}^{a}, A_{\mu}, B_{\mu},$ and $W_{\mu}^{a}$ are
the gauge bosons of $SU(3)_{c}, SU(3)_{new c}, U(1)_{Y}$,
$U(1)_{new Y}$, and $SU(2)_{w}$  respectively \cite{ft2}.
Thus S-Parity is the same as the CP operation
on fermions, and it is parity followed by an exchange operation on bosons.
As a result a fermion and its CP conjugate state have the same charges
upto exchange operation between gauge force groups.
Thus we propose that the
standard model be extended to the gauge group
$G_{SP}:  SU(2)_{w} \times SU(3)_{c}
\times SU(3)_{new C} \times$ $U(1)_{Y} \times U(1)_{new Y} \times S Parity.$
Now, for example, the assignment of the quark doublet to this group is
\begin{equation}
Q_{L} (2;3,\bar{3};1/3,-1/3) \stackrel{CP}{\leftrightarrow}
Q^{c}_{R} (2;\bar{3},3;-1/3,1/3)
\end{equation}
which is more symmetric since the set of quantum numbers for the above  CP
conjugate states is the same upto ordering.

The most interesting consequence of S Parity is that the number of quarks
increase by a factor of 3
(dimensionality of $Q_{L}$ in equation (3) is thrice
the dimensionality in equation (1)).
We can identify this factor of 3, with the
three families if we identify $SU(3)_{new C}$ with a horizontal symmetry
$SU(3)_{H}$. Thus S-Parity leads to an understanding of the triplication of
quark families.  (In the rest of the paper we will use $SU(3)_{H}$ instead
of $SU(3)_{new C}$).
The idea of horizontal symmetries is not new, and has been considered
by several authors \cite{hor,yan,kan,zou}.  However
the choice of the horizontal gauge group
has been largely arbitrary and has ranged from $U(1)_{H}$ to
$SU(3)_{H}$.  Also the assignment of quarks and leptons to these
groups, and the question of whether the horizontal symmetry is vector-like
or not, has been done differently by different authors
\cite{hor,yan,kan,zou}.  The nice thing
about S Parity  is that it not only implies that the horizontal group is
$SU(3)_H$ but also implies its vector nature and that it acts only on the
quarks and not on the leptons.  Thus it is very constraining.

Before we examine the lepton sector, a comment is in order regarding
$U(1)_{new Y}$.
It turns out that symmetrization of the $U(1)$ force can be done without
changing any physics.
The reason is that since $U(1)_Y$ and $U(1)_{new Y}$
are abelian
forces, we can consider linear combinations of them,
namely,
${{Y + Y_{new}} \over {\sqrt{2}}}$
and ${{Y - Y_{new}} \over {\sqrt{2}}}$.
Owing to the fact that for all fermions the
charges $y + y_{new} = 0$ (due to S Parity), this degree of freedom completely
decouples from the fermion sector.  However, depending on the assignment
of $y$ and $y_{new}$ to the Higgs particles in the theory, the new force
may or may not decouple entirely.  In the rest of this paper we will choose
the higgs assignment of $Y_{new}$ such that the combination $y+y_{new} = 0$
for all particles.  With this choice, there is only one non-trivial
$U_(1)$ in  the  theory which is the usual hypercharge for the
standard model.
A posteriori, this is nice since it means that the only force that
really violates
S-parity in the Weinberg-Salam model is $SU(3)$. This is an additional
motivation to symmetrize it as we have done.

Anomaly cancellation \cite{adl}
has been used as an important constraint \cite{mar,moh,foo} in the one
generation standard model to study, for example, electric charge quantization.
Since we have extended the gauge group of the standard model,
it would be interesting to keep the fermion spectrum as
an unknown and
see how far anomaly cancellation determines it. To begin
let us make the following two assumptions.

Assumption A:  $SU(3)_{c}$ and the combination $Q_{em} = I_{3} + {Y \over 2}$
are vector like and $SU(2)_{w}$ is chiral and acts on left handed particles.

Assumption B:  The fermion content of  the theory is such that anomaly
constraints alone (together with assumption A) imply electric
charge quantization.

Now S-Parity and assumption A
imply that $SU(3)_{H}$ is also vector like.  The minimal
non-trivial assignment of {\it all} the gauge fields of $SU(2)_{w} \times
SU(3)_{c} \times SU(3)_{H} \times U(1)_{Y}$
consistent with
assumption A is
$$Q_{L} = (2;3,\bar{3};Y_{Q})$$
$$U_{R} = (1;3,\bar{3};Y_{Q}+1), D_{R} = (1;3,\bar{3},Y_{Q}-1)$$
where we will leave $Y_Q$ arbitrary for now.  We note that at this stage
there are nine $SU(2)$ doublets and hence the Witten anomaly
constraint \cite{wit}
is not satisfied.  Therefore we {\it must} introduce an odd number of
$SU(2)$ doublets.  The assumption of vector-like $Q_{em}$ and chiral $SU(2)$
implies that there must also be $SU(2)$
singlets.
Thus the most general assignment of doublets and singlets,
consistent with assumption A is
\begin{eqnarray}
 L^{i}_{L} = (2;1,1;Y^{i}_{L}),  e^{i}_{R} = (1;1,1;Y^{i}_{L} - 1), \nonumber
\\
\nu^{i}_{R} = (1;1,1;Y^{i}_{L}+1),   \ \ \ \ \  i \le n' \nonumber \\
L^{i}_{L} = (2;1,1;1),  \nu^{i}_{R} \ absent \nonumber \\
e^{i}_{R} = (1;1,1;2), \ \ \ \ \ n' < i \le n'' \nonumber \\
L^{i}_{L} = (2;1,1;-1),  e^{i}_{R} = (1;1,1;-2), \nonumber \\
\nu^{i}_{R} \ absent \ \ \ \ \ n'' < i \le n
\end{eqnarray}
where $Y^{i}_{L}$ is the hypercharge of the $i^{th}$ doublet and n is odd.

Now the $Tr \ SU(2)^{2}U(1)$ anomaly cancellation implies
\begin{equation}
9 Y_{Q} + \sum_{i=1}^{n'} Y^{i}_{L} + \sum_{i=n'+1}^{n''} 1 +
\sum_{i=n''+1}^{n} (-1) = 0
\end{equation}
It is easy  to check that all other triangle anomalies, including the mixed
gravity anomaly,  are automatically satisfied.
Now assumption B implies that $n'=0$, and therfore
\begin{equation}
Y_{Q} = {{n - 2 n''} \over {9}}, \ \ \  \ \ n'' \leq n
\end{equation}
We verify that $Y_{Q}$, (and hence $Q_{em}$), is quantized
since $n$ and $n''$ are integers.  Physically, $n - 2 n''$ is the
number of chiral lepton families!
The values of $n''$ and $n$ that give the observed $Y_{Q} = 1/3$ are
$n = 2n'' + 3$.  Substituting this in equation (4),
it is easy to check that we always have
3 chiral lepton families (with no right-handed neutrino) just
as in the standard model, with a possibility of $n''$ vector-like
lepton families $( n'' \ge 0)$.
Table 1 displays our results for
the fermion content of the theory.

It is interesting that while we can have vector-like lepton families,
additional quark families are not consistent with Assumptions A and B.
Basically what happens is that any additional quark (or colored)
family \cite{ft3} adds to the left-hand
side of equation (5) an unknown hypercharge associated with that family.
All other anomaly constraints will be satisfied due to assumption A, and
hence will not constrain the value of this hypercharge.
We will therefore lose automatic charge quantization and violate assumption
B. Thus absence of any further colored fermions (other than octets) is
a prediction of Assumptions A and B.  On the other hand,
a discovery of a fourth quark family would, according to the ideas of
S-Parity,  imply a corresponding increase in the dimensionality
of the color group - from $SU(3)$ to $SU(N > 3)$.  This would be a
signal to a partial or grand unification of $SU(3)$
with other forces.

Another observation worth making is that assumptions A and
B imply that there is no vector-like anomaly free $U_(1)$ that commutes
with all generators of $SU(2)$ and $SU(3)$. In fact if we demand that
there be no such vector-like U(1), then assumption A
itself will imply
charge quantization.   For theories
with anomaly free
vector-like $U(1)$ (for example $U(1)_{B-L}$), a majorana mass term
for the neutrino can restore
charge quantization as shown in reference \cite{moh}.

We will now move on to the Higgs sector of the theory
\cite{hor,kan,zou}.  We choose the
following representations
of $SU(2) \times SU(3)_{c} \times SU(3)_{H}
\times U(1)_{Y} \times U(1)_{newY}$ based on the criterion of minimality.
\begin{eqnarray}
J_{1}, \ J_{2} = (1;1,3;0,0) \ \
and \ \ K_{1}, \ K_{2} = (1;\bar{3},1;0,0);
\ \ \phi = (2;1,1;1,-1);
\nonumber \\
B_{1}, \ B_{2} = (2;1,8;1,-1) \ \  and \ \ C_{1}, \ C_{2} = (2;8,1;-1,1)
\end{eqnarray}
Since S-parity is parity followed by
an exchange operator on bosons, we will demand that under
S-parity, $J \leftrightarrow K$,  $B \leftrightarrow C$ and $\phi
\rightarrow \tilde{\phi}$
in addition to equation (2).  The absence of
flavour changing neutral currents \cite{yan,kan,har}
suggest that we break $SU(3)_{H}$ at a high
scale.  It requires two $SU(3)_H$ triplets to do this \cite{li},
and so we assume that
$J_{1}$ and $J_{2}$ pick up vacuum expectation values at a scale $M_{H}$.
Note that S-Parity breaks at this scale {\cite{ft4}.
$B_{1}$ and $B_{2}$ are required to give the quarks masses and mixing
angles \cite{zou}.
Several authors have tried to understand fermion masses and mixing angles
using horizontal gauge bosons \cite{hor,yan,kan,zou}.
In the most predictive of these attempts (for example, see reference
\cite{kan})
only one $B$-type Higgs is introduced. Most of the masses and the mixing
angles have to be generated through radiative corrections.
This requires a low scale for horizontal symmetry breaking.
Unfortunately
these efforts have either proved inconsistent with constraints on
flavour changing neutral currents or do not reproduce quark masses
and mixing angles.  However if we assume that the horizontal
symmetry breaking
scale is sufficiently high, then we will require both $B_{1}$ and $B_{2}$.
Now, one combination of  the vacuum expectation values of
$B_{1}$ and $B_{2}$
can be used to give the
up-sector masses, and another linear combination can be
used to give masses to the down sector.  Their relative orientations can
lead to the CKM matrix \cite{kob}.
There are enough variables in the potential that this
will work.  The Higgs $\phi$ is required to give the leptons their masses,
and can also play a role in the quark mass matrix.  It is also worth noting
that we can choose to have explicit CP violating terms since S Parity
does not necessarily imply CP.
We will take up a
detailed analysis of the mass matrices in a future study.
We also refer the reader
to previous  work \cite{hor,kan,zou}
on the quark mass matrices with $SU(3)_{H}$.

Owing to the fact that the Higgs content of the theory gets enriched
by S-parity (see equation (7)), there is the exciting possibility that
the unification of all the standard
model coupling constants is possible using
$G_{SP}$ as an intermediate symmetry.  We recall that
the precision LEP measurements have proven \cite{ama} that
there must be new physics, beyond the standard model, if
the coupling constants are to unify.
To be predictive, we assume
that all coupling constants unify at a scale $M_{U}$ and
that there is only one intermediate scale $M_{H}$ where S-Parity and
$SU(3)_{H}$ break spontaneously. Only the standard model Higgs contributes
to the running of coupling constants between $M_{Z}$ and $M_{H}$.  Beyond
$M_{H}$ all the Higgs in equation (7) contribute,
since the effective theory at this scale is given by the group $G_{SP}$.
All the quark and lepton families contribute from $M_{Z}$ to $M_{U}$, and
the gauge bosons contribute from the appropriate scales onwards.  We
use the usual one-loop renormalization group equation \cite{ama}
and our result is shown in Figure 1.  Inputting the  LEP data,
we find that
$M_{H} = 10^{7} GeV,$ $\alpha^{-1}_{H}(M_{H}) = 22  $,
$M_{U} = 2 \times 10^{15} GeV$ and $\alpha^{-1}_{U} = 30$.
This is very interesting since the above values are consistent both with
current bounds on proton life-time \cite{ama}
and flavour changing neutral currents \cite{yan,kan,har}.  If we
have more than one intermediate scale, or if we change the Higgs content
a bit (for example, make the B-type Higgs lighter and
use color sextets to break $SU(3)_{H}$ instead of
triplets), then it is still possible to achieve the unification,
consistent with experimental bounds, though the scales of $M_{H}$
and $M_{U}$ may change.  The crucial point is that due to S-Parity we
have the C-type and K-type Higgs fields. The contribution of these
fields to the running of the coupling constants is necessary to
achieve  unification consistent with experiments.

In this letter we have suggested that a Strong Parity symmetry
between matter and anti-matter fields
may be at the heart of some of the puzzles of the standard model.

It is a pleasure to thank R.N. Mohapatra, J.C. Pati, K.S. Babu, L. Rana,
S. Kodiyalam and B. Nath for comments and discussions.  This work was
supported in part by NSF Grant 9119745.

\newpage
\vskip0.4in
\begin{tabular}{|c|c|} \hline
Fermions & $SU(2) \times SU(3)_{c} \times SU(3)_{H} \times U(1)_{Y}$\\
\hline
$Q_{L}$ &  $(2,3,\bar{3},1/3) \equiv \left( \begin{array}{ccc}
                                                    u & c & t \\
                                          d & s & b \end{array}
                                                      \right )_{L}$ \\
$U_{R}$ & $(1,3,\bar{3},4/3) \equiv \left( \begin{array}{ccc}
                                                 u & c & t \end{array}
                                                      \right )_{R}$\\
$D_{R}$ & $(1,3,\bar{3},-2/3) \equiv \left( \begin{array}{ccc}
                                                    d & s & b \end{array}
                                                      \right )_{R}$\\ \hline
$i = 1, 2, 3$           &  Three lepton families\\
$L_{L}^{i}$ & $(2,1,1,-1) \equiv \left( \begin{array}{c}
                                             \nu^{i}\\
                                          e^{i}\end{array} \right )_{L}$\\
$e_{R}^{i}$ & $(1,1,1,-2) \equiv   e_{R}^{i}$\\ \hline
$j \ge 0$ & Vector-like leptons may exist\\
$L^{j}_{vec}$ & $(2,1,1,-1) \equiv \left( \begin{array}{c}
                                             \nu^{j}\\
                                          e^{j}\end{array}
                                           \right )_{L,R}$\\
$e^{j}_{vec}$ & $(1,1,1,-2) \equiv   e_{L,R}^{j}$\\ \hline
$Octets$ & Real $SU(3)$ representations may exist\\
\hline
\end{tabular}
\vskip0.4in
Table 1:  The fermion content of the theory, as determined \\
\ \ \ \ \ \ \ \ \ \ \ \ \ \ \ \ \ \ \ \ \ \ \ \ \ \ \
by automatic charge quantization condition.


\begin{thebibliography}{99}
\bibitem{hor}  F. Wilczek and A. Zee, {\it Phys. Rev. Lett.}
{\bf 42}, 421 (1979);
 A. Davidson, M. Koca and K.C. Wali, {\it Phys. Rev. Lett.} {\bf 43}, 92
(1979);
 S. Meshkov and S.P. Rosen, {\it Phys. Rev. Lett.} {\bf 29} 1764 (1972);
 S. Barr and A. Zee, {\it Phys. Rev.} {\bf D 17}, 1854 (1978);
 R. N. Mohapatra, {\it Phys. Rev.} {\bf D 9}, 3461 (1974);
 Y. Chikashige {\it et al.},
{\it Phys. Lett.} {\bf 94B}, 499 (1980)
\bibitem{yan} T. Maehara and T. Yanagida,
{\it Prog. Theor. Phys.} {\bf 60}, 822 (1978); {\bf 61} 1431 (1979);
T. Yanagida, {\it Phys. Rev.} {\bf D 20}, 2986 (1979)
\bibitem{kan} D.R.T. Jones, G.L. Kane and J.P. Leveille,
{\it Nucl. Phys.} {\bf B198}, 45 (1981)
\bibitem{zou} G. Zoupanos, {\it Z. Phys.} {\bf C 11} 27 (1981);
{\it Phys. Lett.} {\bf 115B} 221 (1982);
E. Papantonopoulos and G.Zoupanos, {\it Z. Phys.} {\bf 16} 361 (1983)
\bibitem{pat} K.S. Babu, J.C. Pati and H. Stremnitzer
{\it Phys. Lett.} {\bf B 256},
206 (1991)
\bibitem{fra} P.H. Frampton, {\it Phys. Rev. Lett.}
{\bf 69}, 2889 (1992)
\bibitem{wei} S. Weinberg, {\it Phys. Rev. Lett.}
{\bf 19}, 1264 (1967);
A. Salam, in {\it Proceedings of the VIII Nobel Symposium},
edited by  N. Svartholm (Almpvst and Wiksells, Stockholm, 1968), p. 367;
S.L. Glashow, J. Iliopoulos, and L. Maiani, {\it Phys. Rev.}
{\bf D 2}, 1285 (1970)
\bibitem{geo} H. Georgi,
{\it Lie Algebras in Particle Physics}, Frontiers in Physics Lecture Note
Series, ed. David Pines, (The
Benjamin/Cummings Publishing Company, 1982), p. 80, 166
\bibitem{ft1} $\bar{3} \neq 3$ since the triplet is a complex $SU(3)$
representation
\bibitem{ft2} CP operation on $SU(2)$ doublets will involve the Pauli matrix
$\tau_{2}$ as usual
\bibitem{adl} S. Adler, {\it Phys. Rev.} {\bf 177}, 2426 (1969);
J.S. Bell and R. Jackiw, {\it Nuovo Cimento} {\bf 60A}, 49 (1969);
S. Adler and W. Bardeen, {\it Phys. Rev.} {\bf 182}, 1517 (1969);
H. Georgi and S.L. Glashow, {\it Phys. Rev.} {\bf D 6}, 429 (1972);
C. Bouchiat, J. Iliopoulos, and P. Meyer, {\it Phys. Lett.}
{\bf 38B}, 519 (1972);
D. Gross and R. Jackiw, {\it Phys. Rev.} {\bf D 6}, 477 (1972)
\bibitem{mar} C.Q. Geng and R.E. Marshak, {\it Commun. Nucl.
Part. Phys.} {\bf 18}, 331 (1989);
{\it Phys. Rev.} {\bf D 39}, 693 (1989)
\bibitem{moh} K.S. Babu, and R. N. Mohapatra,
{\it Phys. Rev. Lett}
{\bf 63}, 938 (1989);
{\it Phys. Rev.} {\bf D 41} 271 (1990)
\bibitem{foo} R. Foot {\it et al.},
{\it Mod. Phys. Lett.} {\bf A 5}, 2721 (1990);
 X.G. He, G.C. Joshi, and R.R. Volkas,
{\it Phys. Rev.} {\bf D 41} 278 (1990)
\bibitem{wit} E. Witten, {\it Phys. Lett.} {\bf 117B},
324 (1982)
\bibitem{ft3} Other than real representations like color octets which can be
consistent with A and B.  For e.g., one octet with $Y=0$.
\bibitem{har} R. Cahn and H. Harari, {\it Nucl. Phys.} {\bf B 176},
135 (1980)
\bibitem{li} L. F. Li, {\it Phys. Rev.} {\bf D 9} 1723 (1974)
\bibitem{ft4} We can introduce an S-parity odd singlet to break it at a
higher scale.
\bibitem{kob} M. Kobayashi and K. Maskawa, {\it Prog. Theor. Phys.}
{\bf 49} 652 (1973)
\bibitem{ama} U. Amaldi {\it et al.}, {\it Phys. Lett.}
{\bf B 281} 374 (1992);
J.Ellis, S.Kelly and D.V. Nanapoulos, {\it Phys. Lett.} {\bf B 260},
131 (1991);
U. Amaldi, W. Boer and H. Furstenau, {\it Phys. Lett.} {\bf B 260},
447 (1991);
P. Langacker and M. Luo, {\it Phys. Rev.} {\bf D 44},
817 (1991)
\end{thebibliography}
\end{document}